\documentclass{article}

\usepackage{arxiv}
\usepackage[square,numbers]{natbib}
\usepackage{graphicx}
\usepackage{amsmath}
\usepackage{amssymb}
\usepackage{setspace}

\usepackage[utf8]{inputenc} % allow utf-8 input
\usepackage[T1]{fontenc}    % use 8-bit T1 fonts
\usepackage{hyperref}       % hyperlinks
\usepackage{url}            % simple URL typesetting
\usepackage{booktabs}       % professional-quality tables
\usepackage{amsfonts}       % blackboard math symbols
\usepackage{nicefrac}       % compact symbols for 1/2, etc.
\usepackage{microtype}      % microtypography
\usepackage{lipsum}
\usepackage{graphicx}
\usepackage{array}
\graphicspath{ {./images/} }

\title{Block-diagonalizable two-dimensional generalized Ising systems (BD2DGIS): the free energy}

\author{
 Vadym Sakhno
   \And
 Mykola Sakhno
}

\begin{document}
\maketitle
\begin{abstract}
The paper is a continuation of \cite{sakhno2020matrix}, where the approach was developed to constructing the exact matrix model for any generalized Ising system, and such model was constructed for certain 2d system. In this paper, the properties of the model are specified for light block diagonalization. A corresponding example is considered. For the example, general exact partition function is obtained and analysed. The analysis shows that the free energy does not depend on the amount of rows with a large amount of cells. For the example with light boundary conditions, the partition function is obtained and the specific free energy per spin is plotted.
\end{abstract}

\section{Introduction} \label{sec:Introduction}
In \cite{sakhno2020matrix}, the Ising model was generalized to a system of cells interacting exclusively by presence of shared spins. Within the cells there were interactions of any complexity, the simplest intracell interactions came down to the Ising model. 

\begin{figure} % picture
    \centering
    \includegraphics[width=\textwidth]{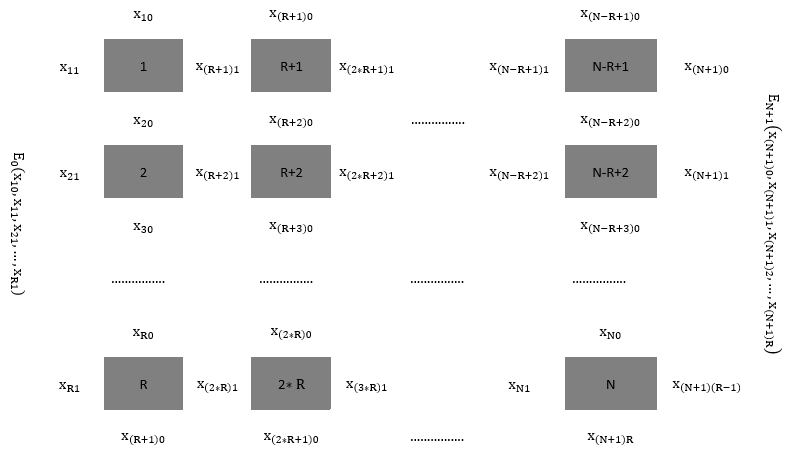}
    \caption{The 2D system under consideration.}
    \label{fig:fig1}
\end{figure}

For simplicity, this paper does not consider local spins, only shared ones, referred to just as spins. The 2D system under consideration is shown in Figure \ref{fig:fig1} (see Figure 1 of \cite{sakhno2020matrix}). It consists of \( (N+2) \) cells having spins with two values \( \{-\frac{1}{2},\frac{1}{2}\} \): $N$ internal cells numbered from $1$ to $N$, start cell numbered $0$, and finish cell numbered $(N+1)$. 

$N$ internal cells form $R$ rows. The lowest spin of a column continues into the highest spin of the next column (see \( x_{\left(R+1\right)0} \) in Figure \ref{fig:fig1}). Thus, cells are placed along a helix. Let the first column be completed, and the last column may be uncompleted. Each internal cell $n$ has four spins, its energy is a given function $E_n\left(x_{n0},\ x_{n1},x_{(n+1)0},x_{(n+R)1}\right)$. Substituting it into (3) of \cite{sakhno2020matrix}, the internal cell function $Z_n\left(x_{n0},\ x_{n1},x_{(n+1)0},x_{(n+R)1}\right)$ is

\begin{equation} \label{eq:1}
        Z_n\left(x_{n0},\ x_{n1},x_{(n+1)0},x_{(n+R)1}\right) = \exp{ \left( -\frac{E_n\left(x_{n0},\ x_{n1},x_{(n+1)0},x_{(n+R)1}\right)}{k_B T} \right) } > 0.
\end{equation}

In \cite{sakhno2020matrix}, for each spin \( x_{n\nu} \in [-\frac{1}{2}, \ +\frac{1}{2}] \), the substituting spin-number \( i_{n\nu} \in [0, 1] \) was introduced according to (6) of \cite{sakhno2020matrix}

\begin{equation} \label{eq:2}
    x_{n\nu}=i_{n\nu}-\frac{1}{2}.
\end{equation}

In the internal cell function (\ref{eq:1}), the substitution of each spin with its spin-number yielded the internal cell frame, which was a set of 16 values, numbered with a compound number of spin-numbers. 
The internal cells in \cite{sakhno2020matrix} may vary, but in this paper they are similar. Therefore, it suffices to consider the first cell $n=1$. Its frame is (17) of \cite{sakhno2020matrix}

\begin{equation} \label{eq:3}
    Z_{1i_{10}i_{11}i_{20}i_{\left(R+1\right)1}}=Z_1\left(i_{10}-\frac{1}{2}, \ i_{11}-\frac{1}{2}, \ i_{20}-\frac{1}{2}, i_{\left(R+1\right)1}-\frac{1}{2}\right).
\end{equation}

 In \cite{sakhno2020matrix}, $2^R\times2^R$ block-diagonal matrices were constructed from the frame, along the diagonal of which there were identical $2\times2$ blocks. Their elements with compound row number  $i_1 i_2 \ldots i_R$ were non-zero only if all row sub-numbers except the last sub-number $R$ were equal to the corresponding column sub-numbers. To emphasize this, these $2^R\times2^R$ block-diagonal matrices were denoted as $2\times2$ matrices with the number $[R]$ (see (25) of \cite{sakhno2020matrix}), for example 

\begin{equation} \label{eq:4}
    \begin{aligned}
    \begin{pmatrix}
        Z_{10000} & Z_{10100} \\
        Z_{10001} & Z_{10101}
    \end{pmatrix}
    _{[R]}
    =
    \begin{pmatrix}
        Z_{10000} & Z_{10100} & 0 & 0 & \ldots & 0 & 0 \\
        Z_{10001} & Z_{10101} & 0 & 0 & \ldots & 0 & 0 \\
        0 & 0 & Z_{10000} & Z_{10100} & \ldots & 0 & 0 \\
        0 & 0 & Z_{10001} & Z_{10101} & \ldots & 0 & 0 \\
        \ldots & \ldots & \ldots & \ldots & \ldots & \ldots & \ldots \\
        0 & 0 & 0 & 0 & \ldots & Z_{10000} & Z_{10100} \\
        0 & 0 & 0 & 0 & \ldots & Z_{10001} & Z_{10101}
    \end{pmatrix}
    .
    \end{aligned}
\end{equation}

Also $2^R\times2^R$ right cyclic shift matrix was introduced in (23) of \cite{sakhno2020matrix}

\begin{equation} \label{eq:5}
\setcounter{MaxMatrixCols}{12}
    \begin{aligned}
        \overleftrightarrow{P_r}
        & =
        \begin{pmatrix}
            1 & 0 & \ldots & 0 & 0 & \ldots & 0 & 0 & \ldots & 0 & 0 & \ldots \\
            0 & 0 & \ldots & 0 & 0 & \ldots & 1 & 0 & \ldots & 0 & 0 & \ldots \\
            0 & 1 & \ldots & 0 & 0 & \ldots & 0 & 0 & \ldots & 0 & 0 & \ldots \\
            0 & 0 & \ldots & 0 & 0 & \ldots & 0 & 1 & \ldots & 0 & 0 & \ldots \\
            \ldots & \ldots & \ldots & \ldots & \ldots & \ldots & \ldots & \ldots & \ldots & \ldots & \ldots & \ldots \\
            \ldots & \ldots & \ldots & \ldots & \ldots & \ldots & \ldots & \ldots & \ldots & \ldots & \ldots & \ldots \\
            0 & 0 & \ldots & 1 & 0 & \ldots & 0 & 0 & \ldots & 0 & 0 & \ldots \\
            0 & 0 & \ldots & 0 & 0 & \ldots & 0 & 0 & \ldots & 1 & 0 & \ldots \\
            0 & 0 & \ldots & 0 & 1 & \ldots & 0 & 0 & \ldots & 0 & 0 & \ldots \\
            0 & 0 & \ldots & 0 & 0 & \ldots & 0 & 0 & \ldots & 0 & 1 & \ldots \\
            \ldots & \ldots & \ldots & \ldots & \ldots & \ldots & \ldots & \ldots & \ldots & \ldots & \ldots & \ldots \\
            \ldots & \ldots & \ldots & \ldots & \ldots & \ldots & \ldots & \ldots & \ldots & \ldots & \ldots & \ldots
        \end{pmatrix}
        .
    \end{aligned}
\end{equation}

Then the sought-for $2^{R+1}\times2^{R+1}$ internal cell matrix was represented in (24) of \cite{sakhno2020matrix} as

\begin{equation} \label{eq:6}
    \overleftrightarrow{Z_1}
    =
    \begin{pmatrix}    
       \begin{pmatrix}
           Z_{10000} & Z_{10100} \\
           Z_{10001} & Z_{10101}
       \end{pmatrix}
       _{[R]}
    &
       \begin{pmatrix}
           Z_{11000} & Z_{11100} \\
           Z_{11001} & Z_{11101}
       \end{pmatrix}
       _{[R]}    
    \\
       \begin{pmatrix}
           Z_{10010} & Z_{10110} \\
           Z_{10011} & Z_{10111}
       \end{pmatrix}
       _{[R]}
    &
       \begin{pmatrix}
           Z_{11010} & Z_{11110} \\
           Z_{11011} & Z_{11111}
       \end{pmatrix}
       _{[R]}    
    \end{pmatrix}
    *
    \begin{pmatrix}
        \overleftrightarrow{P_r} & 0 \\
        0 & \overleftrightarrow{P_r}
    \end{pmatrix}
    ,
\end{equation}

As shown in Figure \ref{fig:fig1}, the start cell energy is $E_0 \left( x_{10}, \ x_{11}, \ x_{21}, \ldots, x_{R1}  \right)$ and the finish cell energy is $E_{N+1} \left(  x_{(N+1)0}, x_{(N+1)1}, x_{(N+1)2}, \ldots, x_{(N+1)R}  \right)$, which can be any given functions of $(R + 1)$ spins. Substituting them into (3) of \cite{sakhno2020matrix} gives the start cell function and the finish cell function (see (\ref{eq:1}) for an internal cell). Then substituting each spin with its spin-number according to (\ref{eq:2}) gives the start cell frame $Z_{0i_{10}i_{11}i_{21} \ldots i_{R1}}$ and the finish cell frame $Z_{(N+1)i_{(N+1)0} \ldots i_{(N+1)(R-1)}i_{(N+1)R}}$ of $2^{R+1}$ values each. The start cell vector $\overleftarrow{Z_0}$ is $2^{R+1}$ column vector of the start cell frame values and the finish cell vector $\overrightarrow{Z_{N+1}}$ is $2^{R+1}$ row vector of the finish cell frame values, let them be called boundary conditions

\begin{equation} \label{eq:7}
   \overleftarrow{Z_0} 
   = \overleftarrow{\|Z_{0i_{10}i_{11}i_{21} \ldots i_{R1}}\|} \ ,
   \quad
   \overrightarrow{Z_{N+1}} 
   = \overrightarrow{\|Z_{(N+1)i_{(N+1)0} \ldots i_{(N+1)(R-1)}i_{(N+1)R}}\|} \ .
\end{equation}

Then the resulting partition function was (9) in \cite{sakhno2020matrix}

\begin{equation} \label{eq:8}
    Z=\overrightarrow{Z_{N+1}} * {\overleftrightarrow{Z_n}}^N * \overleftarrow{Z_0}.
\end{equation}

In this paper, the internal cell matrix (\ref{eq:6}) will be exactly diagonalized for cases of block-diagonalizable 2D generalized Ising systems. It allows finding the free energy in this paper, and various thermodynamic functions in following papers.

\section{BD2DGIS: the definition and an example} \label{sec:Lightly}

\subsection{The definition of BD2DGIS}

The block-diagonalizable will be called 2D generalized Ising systems in which four $2\times2$ matrices composed of (\ref{eq:3}) internal cell frame

\begin{equation} \label{eq:9}
\begin{aligned}
   \begin{pmatrix}
      Z_{10000} & Z_{11000} \\
      Z_{10010} & Z_{11010}
   \end{pmatrix}, \quad
   \begin{pmatrix}
      Z_{10100} & Z_{11100} \\
      Z_{10110} & Z_{11110}
   \end{pmatrix}, \quad
   \begin{pmatrix}
      Z_{10001} & Z_{11001} \\
      Z_{10011} & Z_{11011}
   \end{pmatrix}, \quad
   \begin{pmatrix}
      Z_{10101} & Z_{11101} \\
      Z_{10111} & Z_{11111}
   \end{pmatrix}\\
are \quad diagonalized \quad by \quad the \quad same \quad similarity \quad transformation.
\end{aligned}
\end{equation}

\subsection{An example of BD2DGIS: the internal cell matrix}

\begin{figure}[ht!] % picture
    \centering
    \includegraphics[width=\textwidth]{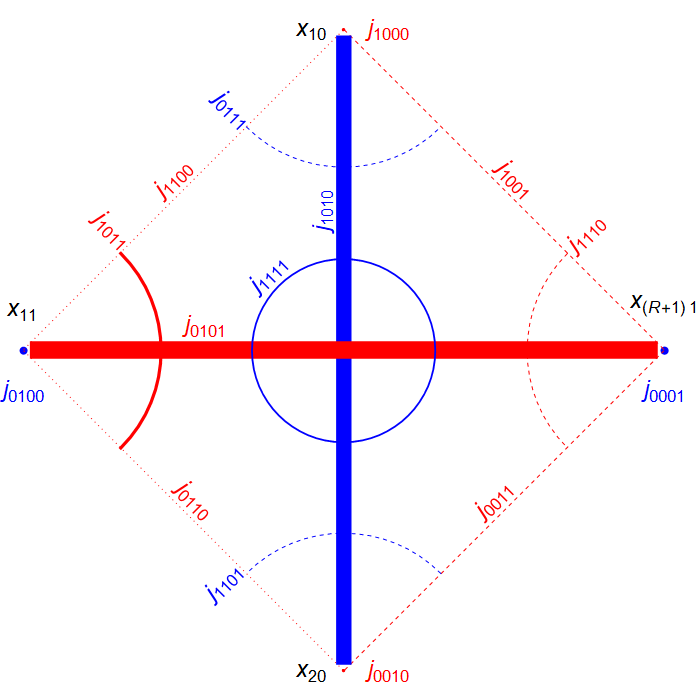}
    \caption{The internal cell parameters of Table \ref{tab:1}.}
    \label{fig:fig2}
\end{figure}

In this article, the system will be considered in which the cell function (\ref{eq:1}) is

\begin{equation} \label{eq:10}
   \begin{aligned}
        &Z_1\left(x_{10},\ x_{11},\ x_{20},\ x_{(1+R)1}\right) = \exp{ \left( -H * \left( j_{1000}*x_{10} + j_{0100}*x_{11} + j_{0010}*x_{20} + j_{0001}*x_{(1+R)1}\right) \right)}
        \\
        *&\exp{ \left( -j_{1100}*x_{10}*x_{11} - j_{0110}*x_{11}*x_{20}-j_{0011}*x_{20}*x_{(1+R)1} - j_{1001}*x_{10}*x_{(1+R)1} \right)}
        \\
        *&\exp{ \left( -j_{1010}*x_{10}*x_{20} - j_{0101}*x_{11}*x_{(1+R)1}-j_{1110}*x_{10}*x_{11}*x_{20} - j_{0111}*x_{11}*x_{20}*x_{(1+R)1} \right)}
        \\
        *&\exp{ \left( -j_{1011}*x_{10}*x_{20}*x_{(1+R)1} - j_{1101}*x_{10}*x_{11}*x_{(1+R)1} - j_{1111}*x_{10}*x_{11}*x_{20}*x_{(1+R)1}  \right)},
   \end{aligned}
\end{equation} 

where the external field $H$ and the internal cell parameters $j$ of (\ref{eq:10}) are given in Table \ref{tab:1} and shown in Figure \ref{fig:fig2}.

\begingroup
\setlength{\tabcolsep}{4pt}
\renewcommand{\arraystretch}{2.6}
\begin{table}[ht!]
\begin{center}
\begin{tabular}{ m{160pt} | m{160pt} } 
\hline
   $H=1$ &
   $j_{1000}=\cfrac{1}{8}\ln{\left(\cfrac{10166}{651}\right)} \approx 0.344$ \\ 
\hline
   $j_{0100}=-\cfrac{1}{8}\ln{\left(\cfrac{31317099}{49504}\right)} \approx -0.806$ & 
   $j_{0010}=\cfrac{1}{8}\ln{\left(\cfrac{10166}{651}\right)} \approx 0.344$ \\
\hline
   $j_{0001}=-\cfrac{1}{8}\ln{\left(\cfrac{4317184}{6279}\right)} \approx -0.817$  & 
   $j_{1100}=\cfrac{1}{4}\ln{\left(\cfrac{20608}{20553}\right)} \approx 0.000668$ \\
\hline
   $j_{0110}=\cfrac{1}{4}\ln{\left(\cfrac{20608}{20553}\right)} \approx 0.000668$ &
   $j_{0011}=\ln{\left(2\sqrt[4]{\cfrac{714}{9269}}\right)} \approx 0.0523$ \\
\hline
   $j_{1001}=\ln{\left(2\sqrt[4]{\cfrac{714}{9269}}\right)} \approx 0.0523$ &
   $j_{1010}=\ln{\left(\cfrac{11}{\sqrt[4]{6618066}}\right)} \approx -1.53$ \\
\hline
   $j_{0101}=\ln{\left(6\sqrt[4]{\cfrac{2346}{2821}}\right)} \approx 1.75$ &
   $j_{1110}=\ln{\left(\cfrac{121}{72}\sqrt{\cfrac{1547}{4278}}\right)} \approx 0.0105$ \\
\hline
   $j_{0111}=\ln{\left(32\sqrt{\cfrac{78}{84847}}\right)} \approx -0.0302$ &
   $j_{1011}=\cfrac{1}{2}\ln{\left(\cfrac{508599}{269824}\right)} \approx 0.317$ \\
\hline
   $j_{1101}=\ln{\left(32\sqrt{\cfrac{78}{84847}}\right)} \approx -0.0302$ & 
   $j_{1111}=-\ln{\left(\cfrac{2821}{2346}\right)} \approx -0.184$ \\
\hline
\end{tabular}
\end{center}
\caption{The external field $H$ and the internal cell parameters $j$ of (\ref{eq:10})}
\label{tab:1}
\end{table}
\endgroup

The internal cell parameters $j$ of (\ref{eq:10}) and Table \ref{tab:1} are displayed in Figure \ref{fig:fig1} as follows:

\begin{itemize}
    \item By the amount of spins in the product with the parameter:
    \begin{itemize}
        \item with one spin - a point in place of the spin (see $j_{1000}$ in $x_{10}$ , $j_{0100}$ in $x_{11}$, $j_{0010}$ in $x_{20}$, $j_{0001}$ in $x_{(1+R)1}$);
        \item with two spins - a line segment between the spins (see $j_{1100}$ between $x_{10}$ and $x_{11}$, $j_{0110}$ between $x_{11}$ and $x_{20}$, $j_{0011}$ between $x_{20}$ and $x_{(1+R)1}$, $j_{1001}$ between $x_{(1+R)1}$ and $x_{10}$, $j_{1010}$ between $x_{10}$ and $x_{20}$, $j_{0101}$) between $x_{11}$ and $x_{(1+R)1}$);
        \item with three spins - an arc around the absent spin (see $j_{0111}$ with $x_{11}*x_{20}*x_{(1+R)1}$ around $x_{10}$, $j_{1011}$ with $x_{10}*x_{20}*x_{(1+R)1}$ around $x_{11}$, $j_{1101}$ with $x_{10}*x_{11}*x_{(1+R)1}$ around $x_{20}$, $j_{1110}$ with $x_{10}*x_{11}*x_{20}$ around $x_{(1+R)1}$);
        \item with four spins - the circle in the center of the cell (see $j_{0111}$ with $x_{10}*x_{11}*x_{20}*x_{(1+R)1}$).
    \end{itemize}
    \item By parameter value:
    \begin{itemize}
        \item more than 0.1 - solid point or line with a thickness of one pixel by 0.1 (in descending order $j_{0101}, j_{1010}, j_{0001}, j_{0100}, j_{1000}, j_{0010}, j_{1011}, j_{1111}$);
        \item more than 0.01 and less than 0.1 - dashed line (in descending order $j_{0011}, j_{1001}, j_{0111}, j_{1101}, j_{1110}$);
        \item less than 0.01 - dotted line ($j_{1100}, j_{0110}$).
    \end{itemize}
    \item By parameter sign:
    \begin{itemize}
        \item plus - red point or line and red name;
        \item minus - blue point or line and blue name.
    \end{itemize}
\end{itemize}

Calculating the frame by (\ref{eq:3}) and substituting it into (\ref{eq:6}) gives the $2^{R+1}\times2^{R+1}$ internal cell matrix

\begin{equation} \label{eq:11}
    \overleftrightarrow{Z_1}
    = \zeta *
    \begin{pmatrix}    
       \begin{pmatrix}
           13 & 69 \\
           68 & 64
       \end{pmatrix}
       _{[R]}
    &
       \begin{pmatrix}
           4 & 22 \\
           24 & 22
       \end{pmatrix}
       _{[R]}    
    \\
       \begin{pmatrix}
           4 & 22 \\
           24 & 22
       \end{pmatrix}
       _{[R]}
    &
       \begin{pmatrix}
           7 & 36 \\
           32 & 31
       \end{pmatrix}
       _{[R]}    
    \end{pmatrix}
    *
    \begin{pmatrix}
        \overleftrightarrow{P_r} & 0 \\
        0 & \overleftrightarrow{P_r}
    \end{pmatrix},
\end{equation}

where

\begin{equation} \label{eq:12}
   \zeta = \cfrac{1}{2*\sqrt[4]{11}*\sqrt[16]{2^{13}*3^5*1103011}} \approx 0.0465.
\end{equation}

Property (\ref{eq:9}) is fulfilled in (\ref{eq:11}).

\subsection{An example of BD2DGIS: light boundary conditions}

In addition to general boundary conditions (\ref{eq:7}), light boundary conditions will be used as follows.

Let spins of the start cell and the finish cell not interact with each other. And the energy of the external field action on each spin $x_{n\nu}$ is the same: $j_0*k_B*T*H*x_{n\nu}$ in the start cell and $j_{N+1}*k_B*T*H*x_{n\nu}$ in the finish cell. Then the start cell function $Z_0$ and the finish cell function $Z_{N+1}$ are

\begin{equation} \label{eq:13}
\begin{aligned}
  &Z_0 \left( x_{10},\ x_{11},\ x_{21},\ldots,x_{R1} \right) =
  \exp{ \left( -H * j_0 * \left( x_{10}+x_{11}+x_{21}+\ldots+x_{R1} \right) \right)} > 0,
  \\
  &Z_{N+1} \left( x_{(N+1)0},x_{(N+1)1},x_{(N+1)2},\ldots,x_{(N+1)R} \right) 
  \\
  & \quad = \exp{ \left( -H * j_{N+1} * \left( x_{(N+1)0}+x_{(N+1)1}+x_{(N+1)2}+\ldots+x_{(N+1)R} \right) \right)} > 0. 
\end{aligned}
\end{equation}

Then substituting each spin 
\( x_{n\nu} \in [-\frac{1}{2}, \ +\frac{1}{2}] \) with its spin-number \( i_{n\nu} \in [0, 1] \) according to (\ref{eq:3}) gives the start cell frame $Z_{0i_{10}i_{11}i_{21} \ldots i_{R1}}$ and the finish cell frame $Z_{(N+1)i_{(N+1)0} \ldots i_{(N+1)(R-1)}i_{(N+1)R}}$ of $2^{R+1}$ values each

\begin{equation} \label{eq:14}
\begin{aligned}
  Z_{0i_{10}i_{11}i_{21} \ldots i_{R1}} 
  & = \prod_{i_{n\nu} \in \{ i_{10}, i_{11}, i_{21}, \ldots, i_{R1} \}} 
  \exp{ \left( -H * j_0 * \left( i_{n\nu} - \frac{1}{2} \right) \right)} > 0,
\\
  Z_{(N+1)i_{(N+1)0}i_{(N+1)1} \ldots i_{(N+1)R}} 
  & = \prod_{i_{n\nu} \in \{ i_{(N+1)0}, i_{(N+1)1}, i_{(N+1)2}, \ldots, i_{(N+1)R} \}} 
  \exp{ \left( -H * j_{N+1} * \left( i_{n\nu} - \frac{1}{2} \right) \right)} > 0,
\end{aligned}
\end{equation}

Now the sought $2^{R+1}$ start cell and finish cell vectors are given by (\ref{eq:7}), taking into account (\ref{eq:14}). However, the internal cell matrix of (\ref{eq:11}) consists of $2^R\times2^R$ blocks. To match them, let $2^R$ column vector $\overleftarrow{1}$ and $2^R$ row vector $\overrightarrow{1}$ of ones be introduced. And the sought $2^{R+1}$ start cell vector $\overleftarrow{Z_0}$ and $2^{R+1}$ finish cell vector $\overrightarrow{Z_{N+1}}$ be rewritten as two identical $2^R$ vectors each

\begin{equation} \label{eq:15}
  \overleftarrow{Z_0} = 
  \begin{pmatrix}
  \overleftrightarrow{\Pi_0} * \overleftarrow{1} \\ 
     \overleftrightarrow{\Pi_0} * \overleftarrow{1}
  \end{pmatrix}, 
  \quad
  \overrightarrow{Z_{N+1}} =\left( \overrightarrow{1} * \overleftrightarrow{\Pi_{N+1}}, \  
     \overrightarrow{1} * \overleftrightarrow{\Pi_{N+1}} \right),
\end{equation}

where

\begin{equation} \label{eq:16}
\begin{aligned}
    \overleftrightarrow{\Pi_0} &= 
      \prod_{\rho = 1}^{R} \overleftrightarrow{Z_0}_{[\rho]} =
      \prod_{\rho = 1}^{R}
        \begin{pmatrix}
            \exp{ \left( \cfrac{H * j_0}{2} \right)} & 0 \\
            0 & \exp{ \left( -\cfrac{H * j_0}{2} \right)}
        \end{pmatrix}_{[\rho]},
    \\
    \overleftrightarrow{\Pi_{N+1}} &= 
      \prod_{\rho = 1}^{R} \overleftrightarrow{Z_{N+1}}_{[\rho]} =
      \prod_{\rho = 1}^{R}
        \begin{pmatrix}
            \exp{ \left( \cfrac{H * j_{N+1}}{2} \right)} & 0 \\
            0 & \exp{ \left( -\cfrac{H * j_{N+1}}{2} \right)}
        \end{pmatrix}_{[\rho]}.
\end{aligned}
\end{equation}

A particular case of (\ref{eq:16}) is the uniform light boundary conditions

\begin{equation} \label{eq:17}
   j_0 = j_{0001} = -\cfrac{1}{8}\ln{\left(\cfrac{4317184}{6279}\right)} \approx -0.817, \quad j_{N+1} = j_{0100} = -\cfrac{1}{8}\ln{\left(\cfrac{31317099}{49504}\right)} \approx -0.806.
\end{equation}

Let conditions (\ref{eq:17}) be considered in detail. In Figure \ref{fig:fig1}, spin $x_{(R+1)1}$ is right for cell $1$ and left for cell $R+1$. According to the first line of (\ref{eq:10}), the energy of the external field action on it is equal to $H* \left( j_{0100} + j_{0001} \right) * x_{(R+1)1}$. Each internal right/left spin has the similar energy. Taking into account (\ref{eq:13}) and (\ref{eq:17}), spins $x_{11},\ x_{21},\ldots,x_{R1}$ of the start cell and spins $x_{(N+1)0},x_{(N+1)1},x_{(N+1)2},\ldots,x_{(N+1)(R-1)}$ of the finish cell have the similar energy.

\section{BD2DGIS: the general exact partition function and its analysis}

\subsection{The block diagonalization of the internal cell matrix}

In the system under consideration, the internal cell matrix (\ref{eq:6}) is (\ref{eq:11}). For it, the matrices of condition (\ref{eq:9}) are

\begin{equation} \label{eq:18}
   \begin{pmatrix}
      13 & 4 \\
      4 & 7
   \end{pmatrix}, \quad
   \begin{pmatrix}
      69 & 22 \\
      22 & 36
   \end{pmatrix}, \quad
   \begin{pmatrix}
      68 & 24 \\
      24 & 32
   \end{pmatrix}, \quad
   \begin{pmatrix}
      64 & 22 \\
      22 & 31
   \end{pmatrix}.\\
\end{equation}

Each matrix of (\ref{eq:18}) is diagonalized by the same matrix $\begin{pmatrix} 2 & -1 \\ 1 & 2 \end{pmatrix}$. Multiplying each element by $2^R\times2^R$ identity matrix $\overleftrightarrow{1}$ gives $2^{R+1}\times2^{R+1}$ matrix

\begin{equation} \label{eq:19}
    \overleftrightarrow{S_0}
    =
    \begin{pmatrix}
      2*\overleftrightarrow{1} & -\overleftrightarrow{1} \\
      \overleftrightarrow{1} & 2*\overleftrightarrow{1}
    \end{pmatrix}
    .
\end{equation}

Let a similarity transformation with matrix $\overleftrightarrow{S_0}$ of (\ref{eq:19}) be performed over the internal cell matrix $\overleftrightarrow{Z_1}$ of (\ref{eq:11})

\begin{equation} \label{eq:20}
    \overleftrightarrow{Z_{1S_0}}
    = {\overleftrightarrow{S_0}}^{-1} * \overleftrightarrow{Z_1} * \overleftrightarrow{S_0} =
    \begin{pmatrix}    
      \overleftrightarrow{B_0}_{[R]} & \overleftrightarrow{0}
    \\
      \overleftrightarrow{0} & \overleftrightarrow{B_1}_{[R]}
    \end{pmatrix}
    *
    \begin{pmatrix}
        \overleftrightarrow{P_r} & 0 \\
        0 & \overleftrightarrow{P_r}
    \end{pmatrix},
\end{equation}

where

\begin{equation} \label{eq:21}
   \overleftrightarrow{B_0}_{[R]} = \zeta *
      \begin{pmatrix}
          15 & 80 \\
          80 & 75
      \end{pmatrix}_{[R]}, \quad
   \overleftrightarrow{B_1}_{[R]} = \zeta *
      \begin{pmatrix}
          5 & 25 \\
          20 & 20
      \end{pmatrix}_{[R]}.
\end{equation}

The resulting matrix $\overleftrightarrow{Z_{1S_0}}$ is the product of two matrices, each of which is block-diagonal of two $2^R\times2^R$ blocks.

\subsection{The general exact partition function}

The partition function is calculated according to (\ref{eq:8}), which may be rewritten as follows

\begin{equation} \label{eq:22}
    Z = \left( \overrightarrow{Z_{N+1}} * \overleftrightarrow{S_0} \right)
      * {\left( {\overleftrightarrow{S_0}}^{-1} 
          * \overleftrightarrow{Z_1} 
          * \overleftrightarrow{S_0} \right)}^N 
          * \left( {\overleftrightarrow{S_0}}^{-1}* \overleftarrow{Z_0} \right)
    = \overrightarrow{Z_{(N+1)S_0}} * {\overleftrightarrow{Z_{1S_0}}}^N * \overleftarrow{Z_{0S_0}}.
\end{equation}

$2^{R+1}$ row vector $\overrightarrow{Z_{(N+1)S_0}}$ can be divided into two halves of $2^R$ row vectors $\left( \overrightarrow{Z_{f0}}, \overrightarrow{Z_{f1}} \right)$ and $2^{R+1}$ column vector $\overleftarrow{Z_{0S_0}}$ can be divided into two halves of $2^R$ column vectors $\begin{pmatrix} \overleftarrow{Z_{s0}}, \\ \overleftarrow{Z_{s1}} \end{pmatrix}$

\begin{equation}\label{eq:23}
    \overrightarrow{Z_{(N+1)S_0}}
    = \overrightarrow{Z_{N+1}} * \overleftrightarrow{S_0}
    = \left( \overrightarrow{Z_{f0}}, \overrightarrow{Z_{f1}} \right),
    \quad
    \overleftarrow{Z_{0S_0}} 
    = {\overleftrightarrow{S_0}}^{-1}* \overleftarrow{Z_0}
    = \begin{pmatrix} \overleftarrow{Z_{s0}}, \\ \overleftarrow{Z_{s1}} \end{pmatrix}.
\end{equation}

Then substitution of (\ref{eq:20}) and (\ref{eq:23}) into (\ref{eq:22}) gives

\begin{equation} \label{eq:24}
    Z = \overrightarrow{Z_{f0}} * {\overleftrightarrow{B_0}_{[R]}}^N * \overleftarrow{Z_{s0}} 
      + \overrightarrow{Z_{f1}} * {\overleftrightarrow{B_1}_{[R]}}^N * \overleftarrow{Z_{s1}}.
\end{equation}

In (26) and (27) of \cite{sakhno2020matrix} some properties of the matrix $\overleftrightarrow{P_r}$ of (\ref{eq:5}) are indicated for any integer $ 1 \leq \rho \leq R $

\begin{equation} \label{eq:25}
    {\overleftrightarrow{P_r}}^R
    =
    \overleftrightarrow{1}
    ,\quad
    \overleftrightarrow{P_r}
    *
    \overleftrightarrow{B}_{[\rho]}
    *
    {\overleftrightarrow{P_r}}^{-1}
    =
    \overleftrightarrow{B}_{[\rho+1]}
    ,\quad
    {\overleftrightarrow{P_r}}^{-1}
    *
    \overleftrightarrow{B}_{[\rho]}
    *
    \overleftrightarrow{P_r}
    =
    \overleftrightarrow{B}_{[\rho-1]}.
\end{equation}

According to Euclid's division lemma, there exist unique integers $N_C$ and $N_0$ such that

\begin{equation} \label{eq:26}
    N = R * N_C + N_0 \quad and \quad 0 \leq N_0 < R,
\end{equation}

where $N$ is the amount of cells, $R$ is the amount of rows, $N_C$ is the amount of completed columns, $N_0$ is the amount of cells in the last incomplete column, while for $N_0=0$ there is no incomplete column.

Then properties (\ref{eq:25}) may be extended

\begin{equation} \label{eq:27}
    {\left( \overleftrightarrow{B}_{[R]} * \overleftrightarrow{P_r}\right)}^{N_0}
    = {\overleftrightarrow{P_r}}^{N_0}
    \ * \prod_{\rho=R-N_0}^{R-1} \overleftrightarrow{B}_{[\rho]},
    \quad
    {\left( \overleftrightarrow{B}_{[R]} * \overleftrightarrow{P_r}\right)}^{R} 
    = \prod_{\rho=1}^{R} \overleftrightarrow{B}_{[\rho]},
\end{equation}

where $\displaystyle \prod_{\rho=\rho_{min}}^{\rho_{max}}$ is equal to 1 if $\rho_{min}>\rho_{max}$ and $\displaystyle \prod_{\rho=1}^{R} \overleftrightarrow{B}_{[\rho]}$ is invariant of $\overleftrightarrow{P_r}$

\begin{equation} \label{eq:28}
      \prod_{\rho=R-N_0}^{R-1} \overleftrightarrow{B}_{[\rho]}
      = 
         \begin{cases} 
            1, & \text{if $N_0 = 0$} \\
            \overleftrightarrow{B}
            _{[R-1]}, & \text{if $N_0 = 1$} \\
            \ldots \\
            \overleftrightarrow{B}
            _{[1]} *
            \overleftrightarrow{B}
            _{[2]} *
            \ldots *
            \overleftrightarrow{B}
            _{[R-1]}, & \text{if $N_0 = R-1$}
         \end{cases}.
\end{equation}

Using (\ref{eq:25}) ... (\ref{eq:28}) in (\ref{eq:24}) gives

\begin{equation} \label{eq:29}
\begin{aligned}
    Z = &\overrightarrow{Z_{f0}}
      * {\overleftrightarrow{P_r}}^{N_0}
      * \prod_{\rho \in[R-N_0,R-1]} {\overleftrightarrow{B_0}_{[\rho]}}^{N_C+1} 
      * \prod_{\rho \notin[R-N_0,R-1]} {\overleftrightarrow{B_0}_{[\rho]}}^{N_C}
      * \overleftarrow{Z_{s0}} \\
    \\
    + &\overrightarrow{Z_{f1}}
      * {\overleftrightarrow{P_r}}^{N_0}
      * \prod_{\rho \in[R-N_0,R-1]} {\overleftrightarrow{B_1}_{[\rho]}}^{N_C+1}
      * \prod_{\rho \notin[R-N_0,R-1]} {\overleftrightarrow{B_1}_{[\rho]}}^{N_C}
      * \overleftarrow{Z_{s1}}.
\end{aligned}
\end{equation}

Let $2^R\times2^R$ matrices be introduced

\begin{equation} \label{eq:30}
    \overleftrightarrow{S_1}_{[\rho]} =
    \begin{pmatrix}
      -3+\sqrt{73} & -3-\sqrt{73} \\
      8 & 8
  \end{pmatrix}
  _{[\rho]}, \quad
    \overleftrightarrow{S_2}_{[\rho]} =
    \begin{pmatrix}
      -3+\sqrt{89} & -3-\sqrt{89} \\
      8 & 8
  \end{pmatrix}
  _{[\rho]}.
\end{equation}

Similarity transformation with matrices $\overleftrightarrow{S_1}_{[\rho]}$ and $\overleftrightarrow{S_2}_{[\rho]}$ of (\ref{eq:30}) over matrices $\overleftrightarrow{B_0}_{[\rho]}$ and $\overleftrightarrow{B_1}_{[\rho]}$ of (\ref{eq:21}) diagonalizes the latter

\begin{equation} \label{eq:31}
\begin{aligned}
   \overleftrightarrow{B_0}_{[\rho]} = &\overleftrightarrow{S_1}_{[\rho]}
      * \left( {\overleftrightarrow{S_1}_{[\rho]}}^{-1} * \overleftrightarrow{B_0}_{[\rho]} 
      * \overleftrightarrow{S_1}_{[\rho]} \right) * {\overleftrightarrow{S_1}_{[\rho]}}^{-1}
      = \lambda_1 * \overleftrightarrow{S_1}_{[\rho]} *
      \begin{pmatrix}
        1 & 0 \\
        0 & \lambda_{21}
      \end{pmatrix}_{[\rho]}
      * {\overleftrightarrow{S_1}_{[\rho]}}^{-1},
     \\
   \overleftrightarrow{B_1}_{[\rho]} = &\overleftrightarrow{S_2}_{[\rho]}
      * \left( {\overleftrightarrow{S_2}_{[\rho]}}^{-1} * \overleftrightarrow{B_1}_{[\rho]} 
      * \overleftrightarrow{S_2}_{[\rho]} \right) * {\overleftrightarrow{S_2}_{[\rho]}}^{-1}
      = \lambda_3 * \overleftrightarrow{S_2}_{[\rho]} *
      \begin{pmatrix}
        1 & 0 \\
        0 & \lambda_{43}
      \end{pmatrix}_{[\rho]}
      * {\overleftrightarrow{S_2}_{[\rho]}}^{-1},
\end{aligned} 
\end{equation}

where taking into account (\ref{eq:12})

\begin{equation} \label{eq:32}
\begin{aligned}
   \lambda_1 = \ &\zeta * \left( 45+10\sqrt{73} \right) \approx 6.063, \quad
   \lambda_{21} = \cfrac{9-2\sqrt{73}}{9+2\sqrt{73}} \approx -0.3100,
   \\
   \lambda_3 = \ &\zeta * \cfrac{25+5\sqrt{89}}{2} \approx 1.677, \quad
   \lambda_{43} = \cfrac{5-\sqrt{89}}{5+\sqrt{89}} \approx -0.3072.
\end{aligned}
\end{equation}

Substituting (\ref{eq:31}) into (\ref{eq:29}) gives the sought general exact partition function

\begin{equation} \label{eq:33}
\begin{aligned}
    Z = &{\lambda_1}^N * \overrightarrow{Z_{f2}}
      * \prod_{\rho \in[R-N_0,R-1]}
         \begin{pmatrix}
           1 & 0 \\
           0 & {\lambda_{21}}^{N_C+1}
         \end{pmatrix}_{[\rho]}
      * \prod_{\rho \notin[R-N_0,R-1]} 
         \begin{pmatrix}
           1 & 0 \\
           0 & {\lambda_{21}}^{N_C}
         \end{pmatrix}_{[\rho]}
      * \overleftarrow{Z_{s2}} \\
    \\
    + &{\lambda_3}^N * \overrightarrow{Z_{f3}}
      * \prod_{\rho \in[R-N_0,R-1]}
         \begin{pmatrix}
           1 & 0 \\
           0 & {\lambda_{43}}^{N_C+1}
         \end{pmatrix}_{[\rho]}
      * \prod_{\rho \notin[R-N_0,R-1]} 
         \begin{pmatrix}
           1 & 0 \\
           0 & {\lambda_{43}}^{N_C}
         \end{pmatrix}_{[\rho]}
      * \overleftarrow{Z_{s3}},
\end{aligned}
\end{equation}

where

\begin{equation} \label{eq:34}
\begin{aligned}
   \overrightarrow{Z_{f2}} = &\overrightarrow{Z_{f0}} *{\overleftrightarrow{P_r}}^{N_0}
      * \prod_{\rho \in[1,R]} \overleftrightarrow{S_1}_{[\rho]},
   \quad
   \quad
   \overleftarrow{Z_{s2}} =  
      \prod_{\rho \in[1,R]} {\overleftrightarrow{S_1}_{[\rho]}}^{-1} * \overleftarrow{Z_{s0}},
    \\
   \overrightarrow{Z_{f3}} = &\overrightarrow{Z_{f1}} *{\overleftrightarrow{P_r}}^{N_0}
      * \prod_{\rho \in[1,R]} \overleftrightarrow{S_2}_{[\rho]},
   \quad
   \quad
   \overleftarrow{Z_{s3}} =  
      \prod_{\rho \in[1,R]} {\overleftrightarrow{S_2}_{[\rho]}}^{-1} * \overleftarrow{Z_{s1}}.
\end{aligned}
\end{equation}

\subsection{Analysis of the general exact partition function} \label{subsec:analysis}

For $N_0=0$, expression (\ref{eq:34}) is analyzed lightly

\begin{equation} \label{eq:35}
    Z = {\lambda_1}^N * \overrightarrow{Z_{f2}}
      * \prod_{\rho=1}^{R}
      \begin{pmatrix}
          1 & 0 \\
          0 & {\lambda_{21}}^{N_C}
      \end{pmatrix}_{[\rho]}
      * \overleftarrow{Z_{s2}}
    + {\lambda_3}^N * \overrightarrow{Z_{f3}}
      * \prod_{\rho=1}^{R}
      \begin{pmatrix}
          1 & 0 \\
          0 & {\lambda_{43}}^{N_C}
      \end{pmatrix}_{[\rho]}
      * \overleftarrow{Z_{s3}} \ .
\end{equation}

Let the first term of (\ref{eq:35}) be analysed in detail. Each $2^R\times2^R$ matrix of the product is strictly diagonal. Let their row/column elements be numbered with compound number having $2^R$ binary sub-numbers $q_1 \ldots q_{\rho} \ldots q_R$. Then index ${[\rho]}$ of any matrix indicates that its elements depend only on $q_{\rho}$, i.e. the element having $q_{\rho}=0$ is equal to 1 and the element having $q_{\rho}=1$ is equal to ${\lambda_{21}}^{N_C}$. Thus, the element $q_1 \ldots q_{\rho} \ldots q_R$ of the product of matrices is equal to ${\lambda_{21}}^{N_C*Q}$ where $Q = q_1 + \ldots + q_{\rho} + \ldots + q_R, \quad 0 \leq Q \leq R$. And the number of the elements with the same $Q$ is equal to the binomial coefficient $\displaystyle \binom{R}{Q} = \frac{R!}{Q!*(R-Q)!}$.

Let $set(Q)$ be the set of numbers having the same $Q$. For example: set(0) has 1 number $000 \ldots 0$, $set(1)$ has $R$ numbers $100 \dots 0, \ 010 \dots 0,$  \textbf{\ldots}  $,\ 000 \dots 1$. Let ${\big \langle \overrightarrow{Z_{f2}} * \overleftarrow{Z_{s2}} \big \rangle}_{set(Q)}$ be the arithmetic mean over $set(Q)$ of the $2^R$ elements of the scalar product $\overrightarrow{Z_{f2}} * \overleftarrow{Z_{s2}}$. Then (\ref{eq:35}) may be written as

\begin{equation} \label{eq:36}
\begin{aligned}
    Z = & {\lambda_1}^N * \sum_{Q=0}^{R} {\lambda_{21}}^{N_C*Q} * \frac{R!}{Q!*(R-Q)!}
    *{\big \langle \overrightarrow{Z_{f2}} * \overleftarrow{Z_{s2}} \big \rangle}_{set(Q)} \\
    \\
    + & {\lambda_3}^N * \sum_{Q=0}^{R} {\lambda_{43}}^{N_C*Q} * \frac{R!}{Q!*(R-Q)!}
    *{\big \langle \overrightarrow{Z_{f3}} * \overleftarrow{Z_{s3}} \big \rangle}_{set(Q)},
\end{aligned}
\end{equation}

or in more detail

\begin{equation} \label{eq:37}
\begin{aligned}
    Z = & {\lambda_1}^N * \left( Z_{f2000 \ldots 0} * Z_{s2000 \ldots 0} 
    + {\lambda_{21}}^{N_C} * R 
    *{\big \langle \overrightarrow{Z_{f2}} * \overleftarrow{Z_{s2}} \big \rangle}_{set(1)} 
    + \ldots \right) \\
    \\
    + & {\lambda_3}^N * \left( Z_{f3000 \ldots 0} * Z_{s3000 \ldots 0} 
    + {\lambda_{43}}^{N_C} * R 
    *{\big \langle \overrightarrow{Z_{f3}} * \overleftarrow{Z_{s3}} \big \rangle}_{set(1)} 
    + \ldots \right).
\end{aligned}
\end{equation}

Let properties $\left| \lambda_{21} \right| < 1, \ \left| \lambda_{43} \right| < 1 \ \text{and} \ \left| \lambda_3 \right| < \lambda_1$ (see (\ref{eq:32})) be taken into account. Then with a large amount of completed columns $N_C$, the partition function of (\ref{eq:37}) is close to

\begin{equation} \label{eq:38}
    Z_{\infty} = {\lambda_1}^N * Z_{f2000 \ldots 0} * Z_{s2000 \ldots 0}. 
\end{equation}

From the partition function one gets the free energy $A$ and the specific free energy per spin $a$

\begin{equation} \label{eq:39}
    A=-k_B*T*ln(Z), \quad a=\frac{A}{2*N+R+1},
\end{equation}

where $2*N+R+1$ is the amount of spins, since each of the $N$ internal cells has 2 spins and the finish cell has $R+1$ spins with spin first sub-number equal to the cell number.

Taking into account (\ref{eq:38}), with a large amount of completed columns $N_C$, the free energy $A$ and the specific free energy per spin $a$ of (\ref{eq:39}) are close to

\begin{equation} \label{eq:40}
    A_{\infty}=-k_B * T * N * ln(\lambda_1), \quad 
    a_{\infty}=-k_B * T * \cfrac{ln(\lambda_1)}{2}.
\end{equation}

Properties of (\ref{eq:40}) are valid not only for simplified expression (\ref{eq:35}), but also for general expression (\ref{eq:33}).

It is important that the properties of (\ref{eq:40}) do not depend on the number of rows $R$ and the boundary conditions.

\section{BD2DGIS with light boundary conditions: the partition function and free energy}

\subsection{The partition function}

Substituting the light boundary conditions of (\ref{eq:15}) in (\ref{eq:23}) with allowance for (\ref{eq:19}) gives

\begin{equation} \label{eq:41}
\begin{aligned}
    \overrightarrow{Z_{(N+1)S_0}} &= 
        \left( \overrightarrow{1} * \overleftrightarrow{\Pi_{N+1}}, \  
        \overrightarrow{1} * \overleftrightarrow{\Pi_{N+1}} \right) * 
        \begin{pmatrix}
          2*\overleftrightarrow{1} & -\overleftrightarrow{1} \\
          \overleftrightarrow{1} & 2*\overleftrightarrow{1}
      \end{pmatrix} 
      = \left( 3 * \overrightarrow{1} * \overleftrightarrow{\Pi_{N+1}}, \  
        \overrightarrow{1} * \overleftrightarrow{\Pi_{N+1}} \right)
      = \left( \overrightarrow{Z_{f0}}, \overrightarrow{Z_{f1}} \right),
   \\
    \overleftarrow{Z_{0S_0}} &= 
      {\begin{pmatrix}
          2*\overleftrightarrow{1} & -\overleftrightarrow{1} \\
          \overleftrightarrow{1} & 2*\overleftrightarrow{1}
      \end{pmatrix}}^{-1}  
      * \begin{pmatrix}
          \overleftrightarrow{\Pi_0} * \overleftarrow{1} \\ 
          \overleftrightarrow{\Pi_0} * \overleftarrow{1}
      \end{pmatrix}
    = \frac{1}{5} * 
    \begin{pmatrix}
      3 * \overleftrightarrow{\Pi_0} * \overleftarrow{1} 
      \\
      \overleftrightarrow{\Pi_0} * \overleftarrow{1}
    \end{pmatrix}
    = \begin{pmatrix} \overleftarrow{Z_{s0}}, \\ \overleftarrow{Z_{s1}} \end{pmatrix}.
\end{aligned}
\end{equation}

Substituting (\ref{eq:41}) in (\ref{eq:34}) with allowance for (\ref{eq:16}) gives

\begin{equation} \label{eq:42}
\begin{aligned}
   \overrightarrow{Z_{f2}} = &3 * \overrightarrow{1} * \prod_{\rho \in[1,R]} 
   \left( \overleftrightarrow{Z_{N+1}}_{[\rho]} * \overleftrightarrow{S_1}_{[\rho]} \right),
   \quad
   \quad
   \overleftarrow{Z_{s2}} = \frac{3}{5} * \prod_{\rho \in[1,R]}
   \left( {\overleftrightarrow{S_1}_{[\rho]}}^{-1} * \overleftrightarrow{Z_0}_{[\rho]} \right) * \overleftarrow{1},
    \\
   \overrightarrow{Z_{f3}} = &\overrightarrow{1} * \prod_{\rho \in[1,R]} 
   \left( \overleftrightarrow{Z_{N+1}}_{[\rho]} * \overleftrightarrow{S_2}_{[\rho]} \right),
   \quad
   \quad
   \quad
   \overleftarrow{Z_{s3}} = \frac{1}{5} * \prod_{\rho \in[1,R]}
   \left( {\overleftrightarrow{S_2}_{[\rho]}}^{-1} * \overleftrightarrow{Z_0}_{[\rho]} \right) * \overleftarrow{1},
\end{aligned}
\end{equation}

where the property $\overrightarrow{1} * \overleftrightarrow{\Pi_{N+1}} * \overleftrightarrow{P_r} = \overrightarrow{1} * \overleftrightarrow{\Pi_{N+1}}$ is taken into account.

Substituting (\ref{eq:42}) in (\ref{eq:33}) gives

\begin{equation} \label{eq:43}
\begin{aligned}
    Z = &\cfrac{9}{5} * {\lambda_1}^N * \overrightarrow{1} 
      * \prod_{\rho \in[R-N_0,R-1]}
      {\overleftrightarrow{D_0}}_{[\rho]} \left( {\lambda_{21}}^{N_C+1} \right)
      * \prod_{\rho \notin[R-N_0,R-1]}
      {\overleftrightarrow{D_0}}_{[\rho]} \left( {\lambda_{21}}^{N_C} \right)
      * \overleftarrow{1} 
      \\
     + &\cfrac{1}{5} * {\lambda_3}^N * \overrightarrow{1} 
      * \prod_{\rho \in[R-N_0,R-1]}
      {\overleftrightarrow{D_1}}_{[\rho]} \left( {\lambda_{43}}^{N_C+1} \right)
      * \prod_{\rho \notin[R-N_0,R-1]}
      {\overleftrightarrow{D_1}}_{[\rho]} \left( {\lambda_{43}}^{N_C} \right)
      * \overleftarrow{1},
\end{aligned}
\end{equation}

where
\begin{equation} \label{eq:44}
\begin{aligned}
   {\overleftrightarrow{D_0}}_{[\rho]} \left( \lambda \right)
   = &\overleftrightarrow{Z_{N+1}}_{[\rho]} * \overleftrightarrow{S_1}_{[\rho]}
         * \begin{pmatrix}
           1 & 0 \\
           0 & \lambda
         \end{pmatrix}_{[\rho]}
   * {\overleftrightarrow{S_1}_{[\rho]}}^{-1} * \overleftrightarrow{Z_0}_{[\rho]},
   \\
   {\overleftrightarrow{D_1}}_{[\rho]} \left( \lambda \right)
   = &\overleftrightarrow{Z_{N+1}}_{[\rho]} * \overleftrightarrow{S_2}_{[\rho]}
         * \begin{pmatrix}
           1 & 0 \\
           0 & \lambda
         \end{pmatrix}_{[\rho]}
   * {\overleftrightarrow{S_2}_{[\rho]}}^{-1} * \overleftrightarrow{Z_0}_{[\rho]}.
\end{aligned}
\end{equation}

Matrices ${\overleftrightarrow{D_0}}_{[\rho]} \left( \lambda \right)$ and ${\overleftrightarrow{D_1}}_{[\rho]} \left( \lambda \right)$  of (\ref{eq:44}) for uniform light boundary conditions can be obtained by substituting (\ref{eq:16}) and (\ref{eq:30}) in (\ref{eq:44}) with allowance for (\ref{eq:17}) and $H=1$

\begin{equation} \label{eq:45}
\begin{aligned}
   {\overleftrightarrow{D_0}}_{[\rho]} \left( \lambda \right)
   = \begin{pmatrix}
     d_{0000} + d_{0001}*\lambda & d_{0010} + d_{0011}*\lambda \\
     d_{0100} + d_{0101}*\lambda & d_{0110} + d_{0111}*\lambda
   \end{pmatrix}_{[\rho]},
   \\
   {\overleftrightarrow{D_1}}_{[\rho]} \left( \lambda \right)
   = \begin{pmatrix}
     d_{1000} + d_{1001}*\lambda & d_{1010} + d_{1011}*\lambda \\
     d_{1100} + d_{1101}*\lambda & d_{1110} + d_{1111}*\lambda
   \end{pmatrix}_{[\rho]},
\end{aligned}
\end{equation}

where coefficients $d$ of (\ref{eq:45}) are given in Table \ref{tab:2}.
\begingroup
\setlength{\tabcolsep}{4pt}
\renewcommand{\arraystretch}{2.7}
\begin{table}[ht!]
\begin{center}
\begin{tabular}{ m{220pt} | m{220pt} } 
\hline
   $d_{0000}=\cfrac{\sqrt{73}-3}{2*\sqrt{73}} * 
   \exp{ \left( H * \cfrac{j_{N+1} + j_0}{2} \right)} \approx 0.144$ &
   $d_{0001}=\cfrac{\sqrt{73}+3}{2*\sqrt{73}} * 
   \exp{ \left( H * \cfrac{j_{N+1} + j_0}{2} \right)} \approx 0.300$ \\
\hline
   $d_{0010}=\cfrac{4}{\sqrt{73}} * 
   \exp{ \left( H * \cfrac{j_{N+1} - j_0}{2} \right)} \approx 0.471$ &
   $d_{0011}=-\cfrac{4}{\sqrt{73}} * 
   \exp{ \left( H * \cfrac{j_{N+1} - j_0}{2} \right)} \approx -0.471$ \\
\hline
   $d_{0100}=\cfrac{4}{\sqrt{73}} * 
   \exp{ \left( H * \cfrac{-j_{N+1} + j_0}{2} \right)} \approx 0.466$ &
   $d_{0101}=-\cfrac{4}{\sqrt{73}} * 
   \exp{ \left( H * \cfrac{-j_{N+1} + j_0}{2} \right)} \approx -0.466$ \\
\hline
   $d_{0110}=\cfrac{\sqrt{73}+3}{2*\sqrt{73}} * 
   \exp{ \left( H * \cfrac{-j_{N+1} - j_0}{2} \right)} \approx 1.521$ &
   $d_{0111}=\cfrac{\sqrt{73}-3}{2*\sqrt{73}} * 
   \exp{ \left( H * \cfrac{-j_{N+1} - j_0}{2} \right)} \approx 0.731$ \\
\hline
   $d_{1000}=\cfrac{\sqrt{89}-3}{2*\sqrt{89}} * 
   \exp{ \left( H * \cfrac{j_{N+1} + j_0}{2} \right)} \approx 0.151$ &
   $d_{1001}=\cfrac{\sqrt{89}+3}{2*\sqrt{89}} * 
   \exp{ \left( H * \cfrac{j_{N+1} + j_0}{2} \right)} \approx 0.293$ \\
\hline
   $d_{1010}=\cfrac{5}{\sqrt{89}} * 
   \exp{ \left( H * \cfrac{j_{N+1} - j_0}{2} \right)} \approx 0.533$ &
   $d_{1011}=-\cfrac{5}{\sqrt{89}} * 
   \exp{ \left( H * \cfrac{j_{N+1} - j_0}{2} \right)} \approx -0.533$ \\
\hline
   $d_{1100}=\cfrac{4}{\sqrt{89}} * 
   \exp{ \left( H * \cfrac{-j_{N+1} + j_0}{2} \right)} \approx 0.422$ &
   $d_{1101}=-\cfrac{4}{\sqrt{89}} * 
   \exp{ \left( H * \cfrac{-j_{N+1} + j_0}{2} \right)} \approx -0.422$ \\
\hline
   $d_{1110}=\cfrac{\sqrt{89}+3}{2*\sqrt{89}} * 
   \exp{ \left( H * \cfrac{-j_{N+1} - j_0}{2} \right)} \approx 1.484$ &
   $d_{1111}=\cfrac{\sqrt{89}-3}{2*\sqrt{89}} * 
   \exp{ \left( H * \cfrac{-j_{N+1} - j_0}{2} \right)} \approx 0.768$ \\
\hline
\end{tabular}
\end{center}
\caption{The coefficients $d$ of (\ref{eq:45})}
\label{tab:2}
\end{table}
\endgroup

In (\ref{eq:43}) the multiplication of the matrices $\overleftrightarrow{D}$ on the left by $\overrightarrow{1}$ and on the right by $\overleftarrow{1}$ gives the sought partition function with uniform light boundary conditions

\begin{equation} \label{eq:46}
\begin{aligned}
    Z = &\cfrac{9}{5} * {\lambda_1}^N 
      * {\left( d_{20} + d_{21} * {\lambda_{21}}^{N_C+1} \right)}^{N_0}
      * {\left( d_{20} + d_{21} * {\lambda_{21}}^{N_C} \right)}^{R-N_0}
      \\
     + &\cfrac{1}{5} * {\lambda_3}^N  
      * {\left( d_{30} + d_{31} * {\lambda_{43}}^{N_C+1} \right)}^{N_0}
      * {\left( d_{30} + d_{31} * {\lambda_{43}}^{N_C} \right)}^{R-N_0}
\end{aligned}
\end{equation}

where coefficients $\lambda$ are given in (\ref{eq:32}) and coefficients $d$ are given in Table \ref{tab:3}.
\begingroup
\setlength{\tabcolsep}{4pt}
\renewcommand{\arraystretch}{2}
\begin{table}[ht!]
\begin{center}
\begin{tabular}{ m{190pt} | m{190pt} } 
\hline
   $d_{20} = d_{0000} + d_{0010} + d_{0100} + d_{0110} \approx 2.601$ &
   $d_{21} = d_{0001} + d_{0011} + d_{0101} + d_{0111} \approx 0.094$ \\
\hline
   $d_{30} = d_{1000} + d_{1010} + d_{1100} + d_{1110} \approx 2.590$ &
   $d_{31} = d_{1001} + d_{1011} + d_{1101} + d_{1111} \approx 0.106$ \\
\hline
\end{tabular}
\end{center}
\caption{The coefficients $d$ of (\ref{eq:46})}
\label{tab:3}
\end{table}
\endgroup

\subsection{The specific free energy per spin}

The specific free energy per spin $a$ of (\ref{eq:39}) may be written as
\begin{equation} \label{eq:47}
    a_T\left( N, R \right) = \cfrac{a}{k_B*T} = -\cfrac{\ln{Z}}{2*N+R+1},
\end{equation}

where the partition function $Z$ is given by (\ref{eq:46}) for uniform light boundary conditions.

Let $a_T$ of (\ref{eq:47}) be plotted versus the amount of cells $N$ for various amounts of rows $R$. 

Let the plot start from the filled first column, that is $N=R$. Then, taking into account (\ref{eq:46}), (\ref{eq:47}) is

\begin{equation} \label{eq:48}
    a_T\left( R, R \right) = -\cfrac{1}{3*R+1} * \ln{ \left( \cfrac{9}{5} * {\lambda_1}^R 
      * {\left( d_{20} + d_{21} * \lambda_{21} \right)}^R
     + \cfrac{1}{5} * {\lambda_3}^R  
      * {\left( d_{30} + d_{31} * \lambda_{43} \right)}^R \right)}.
\end{equation}

With a minimum amount of rows $R=1$, the start of plot (\ref{eq:48}) is

\begin{equation} \label{eq:49}
    a_T\left( 1, 1 \right) = -\cfrac{1}{4} * \ln{ \left( \cfrac{9}{5} * \lambda_1 
      * \left( d_{20} + d_{21} * \lambda_{21} \right)
     + \cfrac{1}{5} * \lambda_3  
      * \left( d_{30} + d_{31} * \lambda_{43} \right) \right)} \approx -0.8412,
\end{equation}

and with a large amount of rows $R \rightarrow \infty$, the start of plot (\ref{eq:48}) is close to

\begin{equation} \label{eq:50}
    a_T\left( \infty, \infty \right) = -\cfrac{1}{3} * \left( \ln{ \lambda_1 } + 
      \ln{ \left( d_{20} + d_{21} * \lambda_{21} \right)} \right) \approx -0.9156.
\end{equation}

Taking into account the property (\ref{eq:40}), the finish of plot (\ref{eq:47}) is close to

\begin{equation} \label{eq:51}
    a_T\left( \infty, R \right) = -\cfrac{\ln{\lambda_1}}{2} \approx -0.9011 
\end{equation}

regardless of the amount of rows $R$.

For plotting, it is convenient to use the amount of columns $N_{C0}$ instead of the amount of cells $N$. Which, given (\ref{eq:26}), is equal to

\begin{equation} \label{eq:52}
    N_{C0} = \cfrac{N}{R} = N_C + \cfrac{N_0}{R}. 
\end{equation}

\begin{figure} % picture
    \centering
    \includegraphics[width=\textwidth]{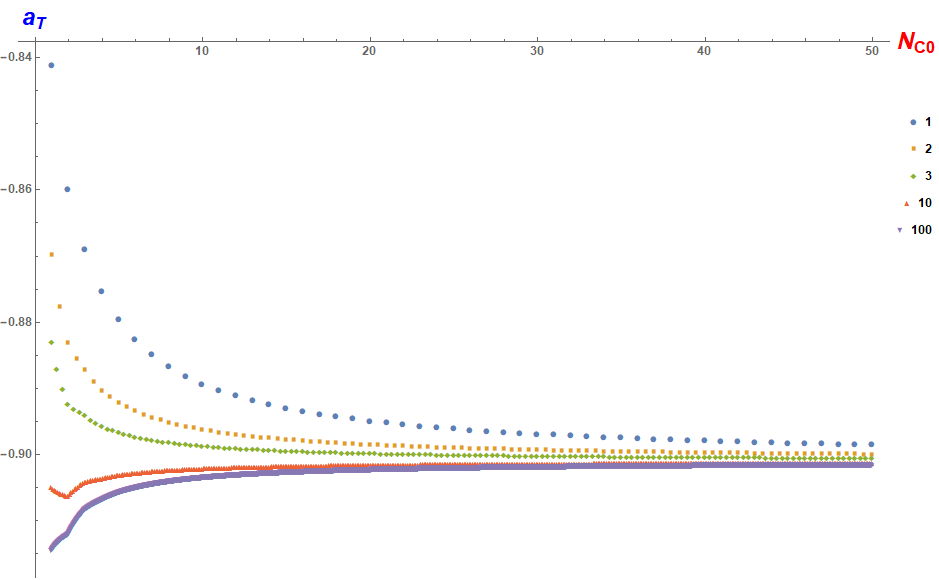}
    \caption{The specific free energy per spin versus the amount of columns for various amounts of rows.}
    \label{fig:fig3}
\end{figure}

Figure \ref{fig:fig3} plots the specific free energy per spin $a_T$ of (\ref{eq:47}) versus the amount of columns $N_{C0} \in [1,50]$ for various amounts of rows $R \in \{1,2,3,10.100\}$. The Wolfram Mathematica microprogram for plotting is given in the Appendix.

\section{Conclusion}

\begin{enumerate}
    \item The properties of internal cell matrix for light block-diagonalization are specified in (\ref{eq:9}).
    \item The example of BD2DGIS is given in Table \ref{tab:1} and Figure \ref{fig:fig2}.
    \item The general exact partition function for the example is obtained in (\ref{eq:33}) and analysed in Subsection \ref{subsec:analysis}.
    \item The analysis showed that the free energy does not depend on the amount of rows with a large amount of cells (see (\ref{eq:40})).
    \item For the example with uniform light boundary conditions, the partition function is obtained in (\ref{eq:46}) and the specific free energy per spin is plotted in Figure \ref{fig:fig3}.
\end{enumerate}

\appendix
\section*{Appendix. The Wolfram Mathematica microprogram for plotting the specific free energy per spin}

\newcommand{\mathsym}[1]{{}}
\newcommand{\unicode}[1]{{}}

\newcounter{mathematicapage}

\begin{doublespace}
\noindent\(\pmb{a[\text{N$\_$},\text{R$\_$}]\text{:=}\text{Module}[\{\text{Nc},\text{N0}\},}\\
\pmb{\{\text{Nc},\text{N0}\}=\text{QuotientRemainder}[N,R];}\\
\pmb{-\text{Log}[1.8*\text{$\lambda $1}{}^{\wedge}N*(\text{d20}+\text{d21}*\text{$\lambda $21}{}^{\wedge}(\text{Nc}+1)){}^{\wedge}\text{N0}*(\text{d20}+\text{d21}*\text{$\lambda
$21}{}^{\wedge}\text{Nc}){}^{\wedge}(R-\text{N0})+}\\
\pmb{0.2*\text{$\lambda $3}{}^{\wedge}N*(\text{d30}+\text{d31}*\text{$\lambda $43}{}^{\wedge}(\text{Nc}+1)){}^{\wedge}\text{N0}*(\text{d30}+\text{d31}*\text{$\lambda
$43}{}^{\wedge}\text{Nc}){}^{\wedge}(R-\text{N0})]/(2*N+R+1)]}\)
\end{doublespace}

\begin{doublespace}
\noindent\(\pmb{\text{Block}[\{\text{$\lambda $1}=6.063,}\\
\pmb{\text{$\lambda $21}=-0.31,}\\
\pmb{\text{$\lambda $3}=1.677,}\\
\pmb{\text{$\lambda $43}=-0.3072,}\\
\pmb{\text{d20}=2.601,}\\
\pmb{\text{d21}=0.094,}\\
\pmb{\text{d30}=2.59,}\\
\pmb{\text{d31}=0.106\},}\\
\pmb{\text{ListPlot}[\text{List}[}\\
\pmb{\text{Legended}\left[\text{Table}\left[\left\{N_{\text{C0}}, a\left[N_{\text{C0}},1\right]\right\},\left\{N_{\text{C0}}, 1 ,50\right\}\right],\text{Placed}[1,\{\{1,1\},\{1,5\}\}]\right],}\\
\pmb{\text{Legended}\left[\text{Table}\left[\left\{N_{\text{C0}}, a\left[2*N_{\text{C0}},2\right]\right\},\left\{N_{\text{C0}} , 1 ,50,0.5\right\}\right],\text{Placed}[2,\{\{1,1\},\{1,6\}\}]\right],}\\
\pmb{\text{Legended}\left[\text{Table}\left[\left\{N_{\text{C0}}, a\left[3*N_{\text{C0}},3\right]\right\},\left\{N_{\text{C0}} , 1 ,50,1/3\right\}\right],\text{Placed}[3,\{\{1,1\},\{1,7\}\}]\right],}\\
\pmb{\text{Legended}\left[\text{Table}\left[\left\{N_{\text{C0}}, a\left[10*N_{\text{C0}},10\right]\right\},\left\{N_{\text{C0}} , 1 , 50,0.1\right\}\right],\text{Placed}[10,\{\{1,1\},\{1,8\}\}]\right],}\\
\pmb{\text{Legended}\left[\text{Table}\left[\left\{N_{\text{C0}}, a\left[100*N_{\text{C0}},100\right]\right\},\left\{N_{\text{C0}} , 1 , 50,0.01\right\}\right],\text{Placed}[100,\{\{1,1\},\{1,9\}\}]\right]],}\\
\pmb{\text{ImageSize}\to \text{Scaled}[0.5],\text{PlotMarkers}\to \{\text{Automatic},10\},}\\
\pmb{\text{PlotRange}\to \text{All},\text{LabelStyle}\to \text{Directive}[\text{Bold},\text{Medium}],}\\
\pmb{\left.\left.\text{AxesLabel}\to \left\{\text{Style}\left[N_{\text{C0}},\text{Large},\text{Bold},\text{Red}\right],\text{Style}\left[a_T,\text{Large},\text{Bold},\text{Blue}\right]\right\}\right]\right]}\)
\end{doublespace}

\bibliographystyle{unsrtnat}
\bibliography{main}

\end{document}